\documentclass[fleqn,10pt]{wlscirep}
\title{The fractal dimensions of Laplacian growth: an analytical approach based on a universal dimensionality function}
\author[1,*]{J. R. Nicol\'as-Carlock}
\author[1]{J. L. Carrillo-Estrada}
\affil[1]{Instituto de F\'isica, Benem\'erita Universidad Aut\'onoma de Puebla, Puebla, 72570, Mexico}
\affil[*]{jnicolas@ifuap.buap.mx}

\begin{abstract}
Laplacian growth, associated to the diffusion-limited aggregation (DLA) model or the more general dielectric-breakdown model (DBM), is a fundamental out-of-equilibrium process that generates structures with characteristic fractal/non-fractal morphologies. However, despite of diverse numerical and theoretical attempts, a data-consistent description of the fractal dimensions of the mass-distributions of these structures has been missing. Here, an analytical description to the fractal dimensions of the DBM and DLA is provided by means of a recently introduced general dimensionality equation for the scaling of clusters undergoing a continuous morphological transition. Particularly, this equation relies on an effective information-function dependent on the Euclidean dimension of the embedding-space and the control parameter of the system. Numerical and theoretical approaches are used in order to determine this information-function for both DLA and DBM. In the latter, a connection to the R\'enyi entropies and generalized dimensions of the cluster is made, showing that DLA could be considered as the point of maximum information-entropy production along the DBM transition. These findings are in good agreement with previous theoretical and numerical results (two- and three-dimensional DBM, and high-dimensional DLA). Notably, the DBM dimensions can be conformed to a universal description independently of the initial cluster-configuration and the embedding-space. 
\end{abstract}

\maketitle

\begin{document}


\section*{Introduction}

The great diversity of fractal morphologies in nature is just matched by the diversity of out-of-equilibrium processes that give origin to them, making the issue of establishing an unified and comprehensive theory of fractal growth a great challenge \cite{benjacob1990,benjacob1997,vicsek1992book,meakin1998book}. In this quest, every so often happens that a simple model emerges to unify diverse phenomena that once seemed to be completely unrelated. This is the case of the Laplacian growth theory, with its emblematic diffusion-limitied aggregation (DLA) model and the more general dielectric breakdown model (DBM), which constitute a paradigm of out-of-equilibrium growth \cite{sander2011}. These models have received significant attention in diverse scientific and technological fields, from the oil industry, through bacterial growth, to cosmology \cite{sander1986,sander2000,sander2011}, even with relevant applications in current neuroscience and cancer research \cite{sturmberg2013,lennon2015,diieva2016}. However, despite of diverse numerical and theoretical attempts at a scaling theory, a data-consistent description of the characteristic fractal dimensions of the DBM and DLA model has been missing \cite{sander2011}. 

In order to have a practical understanding of the problem, let us first recall that the mass-distribution of a given growing structure is can be described in terms of a simple scaling law, $M\propto r^D$, where $r$ is a characteristic length, and $D$ is the dimension of this structure. For example, a sphere growing uniformly in a $d$-dimensional Euclidean space has a dimension, $D=d$, however, in the case of fractal structures, one finds that $D<d$ \cite{vicsek1992book,meakin1998book}. Such is the case of the structures generated by the DLA model, where particles following random-walk trajectories aggregate one-by-one to a seed-particle forming a fractal cluster \cite{sander2011} (see Fig.~1a). This process has been the subject of extensive numerical \cite{ossadnik1991,mandelbrot2002,somfai2003,alves2006,menshutin2012} and theoretical \cite{ball1986,hastings1997,hayakawa1997,davidovitch2000,davidovitch2001,jensen2003,kesten1990,halsey2017} research, not only for the well-known two-dimensional case (where the fractal dimension $D\approx 1.71$ has been commonly reported) but for higher dimensions as well. Although in this case, simulations \cite{meakin1983a,meakin1983b,tolman1989} and theory \cite{ball1984,turkevich1985,turkevich1986,erzan1995,halsey1994,hentschel1984,wang1989,wang1992} are not in best agreement (see Fig.~1b and Table~1). Additionally, there is a still debatable issue regarding the self-similarity of the mass-distribution of the DLA cluster, mostly in two-dimensions. Although there are diverse works that have rigorously proven the consistency of DLA within a self-similar picture, with a fractal dimension very close to $D=1.71$ \cite{ossadnik1991,somfai2003,menshutin2012,davidovitch2001}, there are other results based on multifractal analyses where this is not that clear. In these studies, the DLA cluster is found to be a weak mass-multifractal, that in terms of the generalized dimension, $D_q$ and momenta $q$, goes from $D_{q\to -\infty}\approx1.75$ to $D_{q\to \infty}\approx 1.65$ \cite{vicsek1990,kamer2013}; while in others, it is a monofractal with dimension $D_q\approx 1.7$ for all $q$ \cite{argoul1988,li1989,hanan2012,rodriguez2013}. Nonetheless, among the most important and useful theoretical results found for DLA are: first, that its fractal dimensions are an exclusive function of the embedding Euclidean space, $D=D(d)$ \cite{wang1992}; and secondly, that they must satisfy Kesten's inequality \cite{kesten1990} $D\geq (d+1)/2$ (which is quite restrictive up to $d=3$), and Ball's inequality \cite{ball1984}, $D\geq d-1$, where the equality holds at $d\to\infty$ (which is more restrictive than Kesten for $d\geq 3$) (see Fig.~1b).

Furthermore, DLA is just an instance in more general scenario provided by the DBM. In this generalized model, the growth is related to the growth probability distribution, $\sigma\propto|\nabla\varphi|^\eta$, where $\varphi$ is a scalar field associated to the energy landscape of the growing surface, and $\eta\geq 0$ is the control parameter associated to the net effect of all non-linear interactions influencing the growth \cite{niemeyer1984,hayakawa1986,pietronero1988,amitrano1989,sanchez1993,hastings2001,somfai2004,nagatani1987b,tolman1989b,satpathy1986,vespignani1993} (see Fig.~1c). In two dimensions, for example, by setting $\eta=0$ one makes the growth probability, $\sigma$, proportional to a constant, i.e., one gets a uniform growth probability (or fluctuation independent growth), where the resulting outcome of this process is a compact circular cluster with $D=2$. On the other hand, if $\eta\gg 1$, then, $\sigma$ favours the growth at the tips over the growing front (or fluctuation enhancing growth), where the final outcome of this process is a one-dimensional structure (see Fig.~1c). The most remarkable scenario of this model appears for $\eta=1$, associated to pure stochastic (or fluctuation-preserving) growth dynamics, that corresponds to the universality of the DLA model \cite{sander2011}.

\begin{figure*}[ht!]
\includegraphics[width=\textwidth]{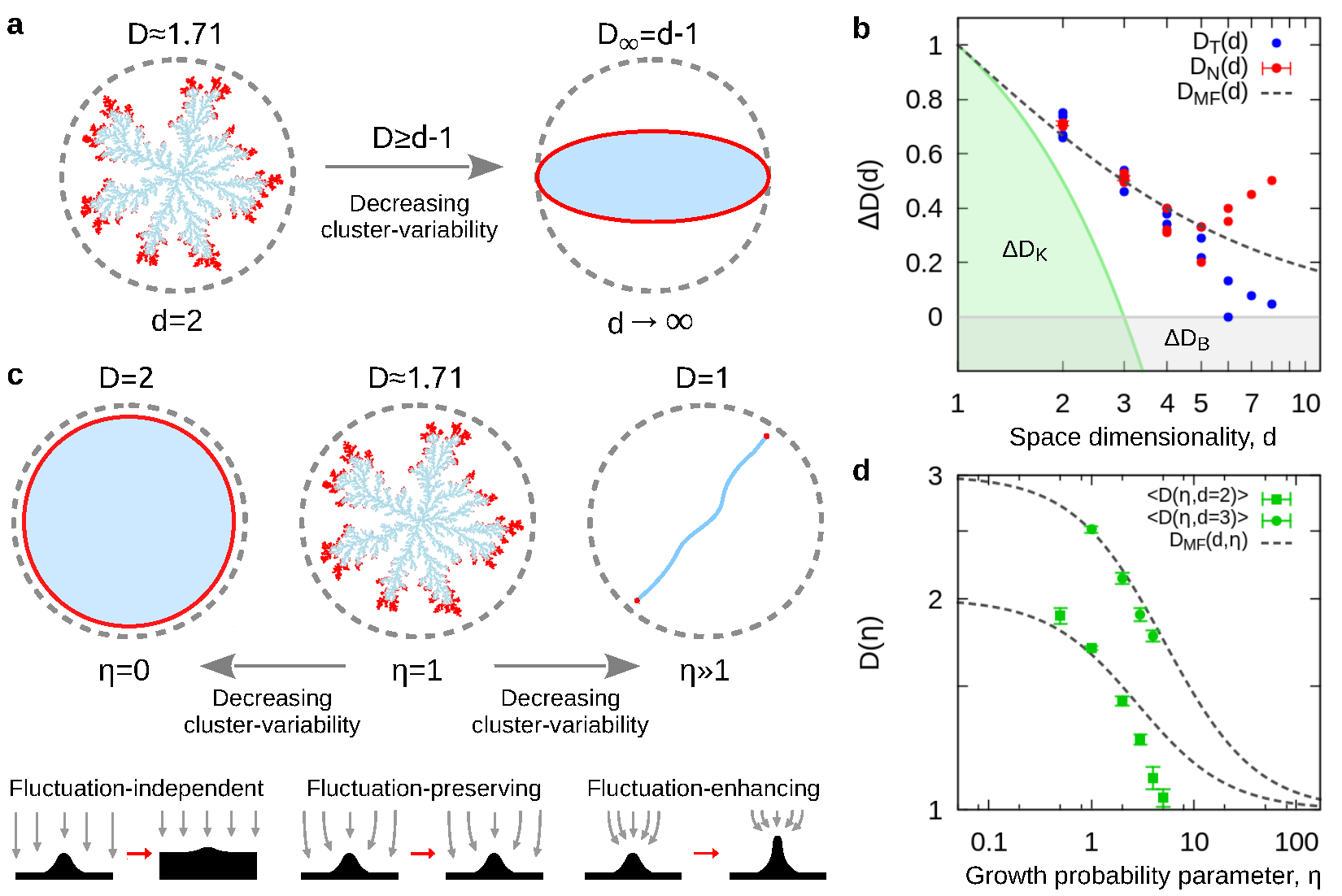}
\caption{\label{fig1} \textbf{The Laplacian framework.} (a) Characteristic features of the DLA fractal (cluster in blue and approximate growing front in red), for $d=2$ and $d\to\infty$. (b) Semi-log plot of the numerical ($D_N$) and theoretical ($D_T$) estimates for $D(d)$ (see Table 1), shown as $\Delta D=D-(d-1)$. Forbidden regions imposed by Kesten (green) and Ball (grey) are indicated as $\Delta D_K$ and $\Delta D_B$, respectively. (c) Characteristic features of the DBM for $d=2$, with a sketch of the corresponding growing dynamics (bottom). (d) Log-log plot of the numerical and theoretical estimates for $D(\eta)$, for $d=2,\,3$ (see Table 2).}
\end{figure*}

The structures generated by the DBM can be characterized by the fractal dimensions $D(d,\eta)$. As function of the parameter $\eta$, they go from isotropic compact structures with $D=d$, when $\eta=0$ (Eden clusters), through intricate dendritic-like fractals with $1<D<d$, for $\eta\approx 1$, to highly anisotropic linear structures, $D\approx 1$, as $\eta\gg 1$. For the two-dimensional case, the collapse to $D=1$ is expected to occur at $\eta\approx 4$ \cite{hastings2001,somfai2004,halsey2002,mathiesen2002,mathiesen2008}. Most characterization approaches rely on numerical methods to estimate $D(\eta)$, mostly in two dimensions, and with very few analytical results for $D(d,\eta)$ \cite{meakin1998book,sander2011}. Among these, mean-field approaches \cite{muthukumar1983,tokuyama1984,matsushita1986} have provided the closed approximation,
\begin{equation}
D_{MF}(d,\eta)=\frac{d^2+\eta}{d+\eta},
\label{eqn:mf}
\end{equation}
that gives a rough description of $D(d,\eta)$, failing to be consistent with the reported numerical results. For example, for DLA in two dimensions, $D_{MF}=5/3\approx 1.67$ differs from the commonly reported $D=1.71$ (see Fig.~1d and Table 2). In general, the derivation of a data-consistent analytical solution to $D(d,\eta)$ has proven to be a non-trivial task and has been missing \cite{meakin1998book,sander2011}. 

In this work, our main goal is to provide a data-consistent analytical description to the fractal dimensions, $D(d,\eta)$, of the DBM and DLA model. To this end, we will make use of a recently introduced framework for the study fractal/non-fractal morphological transitions \cite{nicolas2016,nicolas2017}. In this framework, a morphological transition is defined as the geometrical transformation that a given structure undergoes as a result of the stochastic/energetic (or symmetry-breaking) aspects of its growth-dynamics \cite{nicolas2016}. Quantitatively, the different geometrical features displayed by these structures are described through the scaling or fractal dimension of their mass-distribution, $D$. In addition, all the information regarding symmetry-breaking effects are encoded into an effective control or information-function, $\Gamma$. The fractal dimensions that characterize the transition are given by the dimensionality function\cite{nicolas2017},
\begin{equation}	
D(D_0,\Gamma)=1+(D_0-1)e^{-\Gamma},
\label{eqn:dphi}
\end{equation}
where $D_0$ is the dimension of the initial cluster configuration, and where the functional form of $\Gamma$ is to be found according to the particular phenomenology of the system. Hence, in this framework, finding the complete solution to $D$ implies finding the corresponding information-function, $\Gamma$, of the system.

In the following, we present the results of using equation~(\ref{eqn:dphi}), along with other theoretical and numerical results, in order to find the desired general solution to the fractal dimensions of the DBM and DLA model, $D(d,\eta)$. Without loss of generality and for all practical purposes, we will consider that the mass-distribution of both DBM and DLA clusters is self-similar, that is, it can be defined by a single scaling or fractal dimension \cite{sander2011} (more on this in the Discussion section). Under these considerations, the results are presented as follows: first, we present the conditions that the information-function, $\Gamma$, must satisfy in order to describe the DBM dimensions, as well as a proposal for its general functional form. Secondly, based on this functional form, we present the results of a first numerical approach to determine $\Gamma$ for $d=2$ and $d=3$. Thirdly, we will restrict our attention to the solution of the DLA dimensions for any $d$. Finally, we focus on the general solution to the DBM dimensions where attention is be paid to a theoretical approach that connects the information-entropy of the clusters to their fractal dimension. This result will be tested for $d=2$ and $d=3$. Lastly, we present the discussion of these results and some final remarks. 


\section*{Results}

\subsection*{Conditions for a general $\Gamma$}

Insight into the general form of the information-function, $\Gamma(d,\eta)$, can be gained from the analysis of mean-field equation, $D_{MF}(d,\eta)$, and the reported numerical estimates of $D(d,\eta)$ in two and three dimensions.

First, the mean-field result given in equation~(\ref{eqn:mf}) belongs to a special case of equation~(\ref{eqn:dphi}). This is observed by expanding the exponential of equation~(\ref{eqn:dphi}) up to its first-order term in $\Gamma$, leading to, $D^{(1)}(\Gamma)=1+(D_0-1)/(1+\Gamma)=(D_0+\Gamma)/(1+\Gamma)$. From direct comparison to equation~(\ref{eqn:mf}), one observes that, $D^{(1)}=D_{MF}$, with $D_0=d$, and $\Gamma_{MF}(d,\eta)=\eta/d$. Even though this mean-field result does not provide the correct description to $D(d,\eta)$, this example serves two purposes: it provides a first glimpse into the relation between the information-function and the specific variables of the model, and it provides useful evidence about the validity and generality of equation~(\ref{eqn:dphi}) as a fractality function for the DBM. 

\begin{figure*}[ht!]
\includegraphics[width=\textwidth]{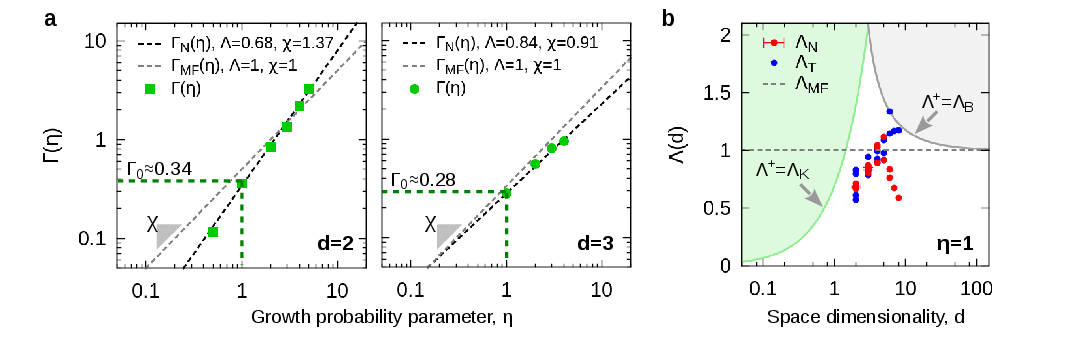}
\caption{\label{fig2} \textbf{Information function.} (a) Log-log plot of $\Gamma(d)$ as provided by the numerical average $\langle D(\eta)\rangle$ in $d=2$ and $d=3$ (in Table 2), the mean-field approximation, $\Gamma_{MF}(\eta)=\eta/d$, and fitting-function, $\Gamma_N(\eta)=(\Lambda/d)\eta^\chi$. (b) Semi-log plot of $\Lambda(d)$ as provided by numerical ($\Lambda_N$) and theoretical ($\Lambda_T$) estimates for $D(d)$ (in Table 1). Forbidden regions imposed by Kesten ($\Lambda_K$) and Ball ($\Lambda_B$), are indicated accordingly.}
\end{figure*}

Secondly, the general form of $\Gamma(d,\eta)$ can be inferred from the data for $D(d,\eta)$ in two and three dimensions. This is done by solving for $\Gamma(D_0,\eta)$ in equation~(\ref{eqn:dphi}), which gives,
\begin{equation}
\label{eqn:gamma} 
\Gamma(D_0,\eta)=-\log D^*,
\end{equation}
where, $D^*=(D-1)(D_0-1)$, is defined as the normalized dimension. As it can be easily seen, $D^*=1$ when $D=D_0$, and $D^*=0$, when $D=1$. Furthermore, considering that for the DBM, $D_0=d$, equation~(\ref{eqn:gamma}) can be used along with the numerical data for $D(d,\eta)$ in Table 2 to obtain a qualitative description of $\Gamma(d,\eta)$ (see Fig.~2a). In these log-log plots, the data suggests that the functional form of $\Gamma$ must be quite close to a power-law relation. Therefore, we propose the following \emph{ansatz}, 
\begin{equation}
\Gamma(d,\eta)=\Gamma_0\eta^\chi,
\label{eqn:gamma_ansatz} 
\end{equation}
with $\Gamma_0=\Lambda/d$, and where $\Lambda$ and $\chi$ are two characteristic real numbers associated to the particular growth dynamics of the system. For example, for the mean-field approximation, $\Gamma_{MF}=\eta/d$, we have the exact solutions $\Lambda_{MF}=1$ and $\chi_{MF}=1$, for all $d$. In general, the specific values of $\Lambda$ and $\chi$ must be determined accordingly.

\subsection*{Numerical approach to DBM: finding a general $\Gamma(d,\eta)$}

One direct way to obtain the desired $\Lambda$ and $\chi$, is by using $\Gamma(d,\eta)$ in equation~(\ref{eqn:gamma_ansatz}) as a fitting-function. To do this, let us first observe that according to equation~(\ref{eqn:gamma_ansatz}), for a fixed $d$-dimensional space, $\Gamma(d,\eta)\to\Gamma(\eta)\propto \eta^\chi$ provides all the information associated to the DBM dimensions; whereas for $\eta=1$, we have that $\Gamma(d,\eta)\to\Gamma_0(d)=\Lambda/d$ is associated to the DLA dimensions. This suggests that the values for $\Lambda$ can be estimated by a simply substitution from the previous knowledge of the fractal dimensions of DLA, while the values for $\chi$ can be obtained via linear-fitting (see Fig.~2a). For example, by substituting $\eta=1$ and $D=1.71$, in equations~(\ref{eqn:gamma}) and (\ref{eqn:gamma_ansatz}), with $D_0=d=2$, we have that $\Lambda=-d\log(D^*)\approx 0.68$. Now, using equation~(\ref{eqn:gamma_ansatz}) as a fitting-function over the data for $\Gamma(\eta)$, we have that $\chi=1.37\pm 0.02$. Following the same procedure for $d=3$, we have that $\Lambda\approx 0.84$ (using $D=2.51$ for three-dimensional DLA) and $\chi=0.91\pm 0.02$, via linear-fitting (see Fig. 2a). 

Although this procedure on its own provides a complete data-consistent description to $D(d,\eta)$, for $d=2$ and $d=3$ (see Table 2), it is only useful when the dimensions of DLA and DBM are known beforehand, that is, it as a good characterization scheme. Nevertheless, this example serves two purposes: first, it shows that in order to find the fractal dimensions of DLA, one must determine the values of $\Lambda=\Lambda(d)$; secondly, once $\Lambda$ is determined, the general solution to the fractal dimensions of the DBM relies on finding the corresponding values of $\chi$. In the following sections we present two methods to determine general expressions for these two quantities.

\subsection*{Solution to DLA: finding a general $\Lambda(d)$}

The purpose of this section is to show how to determine a general expression for $\Lambda(d)$, which can then be used along with equations~(\ref{eqn:dphi}) and (\ref{eqn:gamma_ansatz}), with $\eta=1$, as a solution to the fractal dimensions of DLA. 

Some insight into this problem can be gained from previous numerical and theoretical estimates for $D(d)$ and other rigorous theoretical results. In particular, the data in Table 1 can be used together with $\Lambda(d)=-d\log(D^*)$ to have look into the qualitative behaviour of $\Lambda(d)$ (see Fig.~2b). As expected, $\Lambda(d)$ does not conform to the trivial mean-field approximation, $\Lambda_{MF}=1$. Additionally, some theoretical restrictions can be established. For example, the Kesten's inequality, $D\geq D_K=(d+1)/2$ \cite{kesten1990}, and the Ball's inequality, $D\geq D_B=d-1$ \cite{ball1984}, impose an strict upper boundary, $\Lambda^{+}$, which is defined by parts as $\Lambda^+=\Lambda_K=d\log 2$, for $d\leq 3$, and $\Lambda^+=\Lambda_B=-d\log[(d-2)/(d-1)]$, for $d\geq 3$. From these boundaries, any solution must satisfy $\Lambda\leq\Lambda^+$, where the equality, $\Lambda=\Lambda_B$, holds for $d\to\infty$. In fact, it is only in this limit that $\Lambda_{MF}=\Lambda_B=1$ (see Fig.~2b).

\begin{figure*}[ht!]
\includegraphics[width=\textwidth]{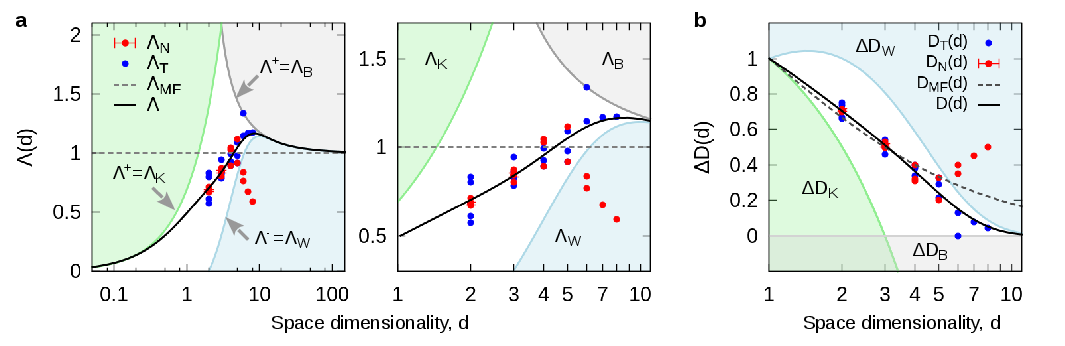}
\caption{\label{fig3} \textbf{The DLA solution.} (a) Semi-log plot of $\Lambda(d)$ with its corresponding analytical solution. (b) Semi-log plot of $D(d)$, shown as $\Delta D=D-(d-1)$. In both (a)-(b) plots, the forbidden regions imposed by Kesten ($\Lambda_K$), Ball ($\Lambda_B$), and Wang ($\Lambda_W$), are indicated accordingly. For the numerical values of $D(d)$, see Table 1.}
\end{figure*}

Considering the previous results, an analytical expression for $\Lambda(d)$ is constructed by using a particular real-space renormalization-group (RG) result for the fractal dimensions of \textsl{on-lattice} DLA \cite{wang1989,wang1992}. Under this RG approach, the DLA dimensions are given in discrete form by, 
\begin{align}
\label{eqn:rg}
D(k,d)&=1+\log \mu/\log 2,\\
\mu(k,d)&=1 + 2\binom{d-1}{1}\phi_1 + \sum_{k=2}^{d-1}\binom{d-1}{k}\phi_k, \nonumber
\end{align}
where, $\mu=\mu(k,d)$ is inversely proportional to the maximum growth probability, $\sigma(d)_{max}=\mu^{-1}$, and where $\phi_k=\phi_k(d)$ are growth potentials that need to be determined for a given lattice configuration. For example, for the square lattice, these equations lead to $D\approx 1.74$, and for a cubic lattice, they lead to $D\approx 2.52$. Independently of the lattice configuration, $\phi_\infty=1/2$ and $\mu_\infty=2^{d-2}+d/2$ are found in the $d\to\infty$ limit \cite{wang1992}. 

For our main task, these \textsl{lattice-dependent} results can be extended to the \textsl{lattice-independent} scenario by replacing the discrete lattice potentials, $\phi_k(d)$, with a continuous effective potential, $\Phi(d)$, i.e., $\phi_k(d)\to\Phi(d)$. Now, for continuity and clarity purposes, we invite the reader to consult the Methods section for further details on how this extension is specifically done. After the corresponding analysis, we found that the \textsl{off-lattice} DLA dimensions, under the RG model, are given by,
\begin{align}
\label{eqn:rgx}
D(d)&=1+\log\mu/\log 2, \\ 
\mu(d)&=1+2(\mu_\infty-1)\Phi, \nonumber \\
\mu_\infty(d) &=2^{d-2}+d/2, \nonumber \\
\Phi(d)&=\Phi_\infty/[1+2^{-r(d-1)}], \nonumber
\end{align}
where $\Phi_\infty=1/2$, and $r=0.762\pm 0.014$. 

By substituting $D(d)$ of equations~(\ref{eqn:rgx}) into equation~(\ref{eqn:gamma}), and solving for $\Lambda$ using equation~(\ref{eqn:gamma_ansatz}), we have that,
\begin{equation}
\label{eqn:gamma_solution}
\Lambda(d)=d\Gamma_0=-d\log(\log\mu/\log 2^{d-1}).
\end{equation}

As shown in Fig.~3a, this analytical expression for $\Lambda$ is in great agreement with the KBW-restrictions, namely, $\Lambda\leq\Lambda_K=d\log 2$ (Kesten's condition), $\Lambda\leq\Lambda_B=-d\log[(d-2)/(d-1)]$ (Ball's condition), and the additional lower boundary, $\Lambda\geq\Lambda^-=\Lambda_W=-d\log[\log\mu_\infty/\log 2^{d-1}]$ (Wang's condition), that come from the inequalities summarized below, 
\begin{align}
\label{eqn:kbw_conditions}
D(d\leq 3)\geq D_K &=(d+1)/2, \\
D(d\geq 3)\geq D_B &=d-1, \nonumber \\
D(d\geq 1)\leq D_W &=1+\log\mu_\infty/\log 2,\nonumber
\end{align}
where $\mu_\infty$ is given in equations~(\ref{eqn:rgx}), and the equality, $D=D_W=D_B$, holds for the $d\to\infty$ limit \cite{wang1992} (see Fig.~3a). Furthermore, the analytical solution provided by equations~(\ref{eqn:rgx}) to the DLA dimensions not only is in great agreement with the KBW-restrictions but provide an accurate description of the data (see Fig.~3b). The numerical values obtained for $D(d)$ are shown in Table 1.

\begin{figure*}[ht!]
\includegraphics[width=\textwidth]{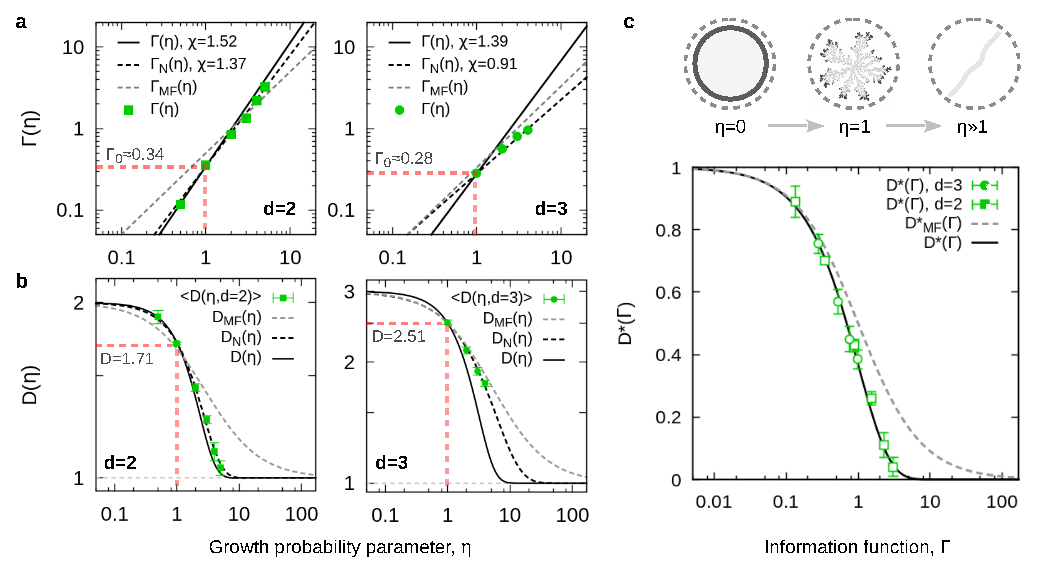}
\caption{\label{fig4} \textbf{The DBM solutions.} (a) Log-log plots of $\Gamma(\eta)$, and (b) semi-log plots of $D(\eta)$, with their corresponding numerical ($\Gamma_N$, $D_N$) and theoretical ($\Gamma$, $D$) solution for $d=2$ and $d=3$. For the numerical values of $D(d,\eta)$ see Table 2. (c) Universal behaviour of the normalized dimension, $D^*(\Gamma)$, with a sketch of the DBM morphological transition for $d=2$, on top.}
\end{figure*}

\subsection*{Solution to DBM: finding a general $\chi(d)$}

Once a general solution to $\Lambda(d)$ has been found, here we propose a theoretical approach to determine $\chi(d)$. This approach is based on a particular finding from the study of the rate of information-entropy production of random fractals \cite{kaufman1989a,kaufman1989b}. This finding suggests that DLA can be associated to a critical state that defines the point of maximum information-entropy production along the DBM morphological transition \cite{dimino1989c,kaufman1989d,kaufman1989e,gleiser2015}. Then, if this is indeed the case, this could manifest itself in the fractality of the system. 

This observation is incorporated into our model for $D(d,\Gamma)$ by means of the formalism of multifractal sets as applied to the mass-distribution of fractal clusters \cite{sander2011,vicsek1990}. In particular, through the relation between the R\'enyi entropies, $S_q$, and the generalized dimension, $D_q$ \cite{zmeskal2013,kamer2013,rodriguez2013}. First, the R\'enyi entropies are defined as, $S_{q}=\log\sum_{i=1}^n p_i(\epsilon)^q/(q-1)$, where $q$ are the momenta, $n$ is the number of partitions (boxes) of the set, and $p_i(\epsilon)$ is the probability of finding an element of the cluster at a spatial observation scale, $\epsilon$. Secondly, the generalized dimension, $D_q$, is related to the R\'enyi entropies by, $D_{q}=\lim_{\epsilon\to 0} S_{q}/\log\epsilon$, in such a way that, for example, for $q=0$, one has the box-counting dimension $D_{q=0}$, for $q=2$, the correlation dimension $D_{q=2}$, and for $q=1$, one has the information-dimension $D_{q=1}$, obtained from the scaling of the information-entropy $S_{q=1}=-\sum_{i=1}^n p_i(\epsilon)\log p_i(\epsilon)$ \cite{sander2011,zmeskal2013}. In the case of self-similar clusters, the generalized dimension, $D_q$, becomes $q$-independent, making all the dimensions $D_q$ equivalent ($D_q\to D$), and directly proportional to the information-entropy ($S_{q=1}\to S$). Then, from the definition of the generalized dimension, we simply have that $S=kD$, with $k=\log(\epsilon)$ \cite{zmeskal2013}. 

This relation implies that $\partial D/\partial\eta=k^{-1}\partial S/\partial\eta\to 0$, in both of the limits $\eta\to 0$ or $\eta\gg 1$. For example, the amount of information needed to characterize a (non-fractal) compact circular or spherical cluster ($D\to d$) as $\eta\to 0$, or the (non-fractal) linear structure ($D\to 1$) for $\eta\gg 1$, does not grow as much as the one needed to characterize the intermediate (disordered) fractal clusters for $\eta\approx 1$ (see Fig. 1b). Therefore, if the DLA fractal corresponds to a maximum in information-entropy production in the DBM transition, this should manifest in the fractality of the cluster itself, specifically, it must show up at a certain point, $\eta_i$, where $\partial S/\partial \eta|_{\eta=\eta_i}$ becomes a global maximum, that is, $\partial^2 S/\partial \eta^2=k\partial^2 D/\partial \eta^2|_{\eta=\eta_i}=0$. 

From equations~(\ref{eqn:dphi}) and (\ref{eqn:gamma_ansatz}), this inflection point, $\eta_i$, satisfies, $[(\partial\Gamma/\partial\eta)^2-\partial^2\Gamma/\partial\eta^2]|_{\eta=\eta_i}=0$, leading to, 
\begin{equation}
\Gamma_0\eta_i^\chi=\frac{\chi-1}{\chi}.
\label{eqn:etai} 
\end{equation}

Finally, by using the DLA condition, $\eta_i=1$, and solving for $\chi$, we have that $\chi=1/(1-\Gamma_0)$, where $\Gamma_0=\Lambda/d$. In this way, the values of $\chi$ depend on $\Lambda$, and the information-function is now given by,
\begin{equation}
\Gamma(d,\eta)=\Gamma_0\eta^{1/(1-\Gamma_0)}
\label{eqn:gamma_theory} 
\end{equation}
where $\Gamma_0=\Lambda/d$ is already given by equation~(\ref{eqn:gamma_solution}). 

Using this result, we have that in two dimensions, $\Lambda\approx 0.68$, and $\chi=1/(1-\Lambda/d)\approx 1.52$, with $D\approx 1.71$ at $\eta=1$; in three dimensions, $\Lambda\approx 0.84$, and $\chi\approx 1.39$, with $D\approx 2.51$ at $\eta=1$ (see Figs. 4a and 4b). The complete numerical values obtained for $D(d,\eta)$ under this approach are shown in Table 2.

\section*{Discussion}

Despite the morphological complexity of the DBM and DLA clusters, equation~(\ref{eqn:dphi}) can be used to describe the fractal dimensions of their mass-distribution, $D(d,\Gamma)$, by considering that the fundamental elements of its growth dynamics are encoded in an effective information-function, $\Gamma(d,\eta)$, as proposed within the framework for fractal to non-fractal morphological transitions. In particular, based on the DBM phenomenology, equation~(\ref{eqn:gamma_ansatz}) is presented as an \textsl{ansatz} to this information-function.

\textbf{\textsl{Solution to DLA.}} From equations~(\ref{eqn:dphi}) and (\ref{eqn:gamma_ansatz}), the DLA dimensions correspond to $\eta=1$, which is $D(d,\eta)\to D(d)$, and $\Gamma(d,\eta)\to \Gamma_0=d^{-1}\Lambda$. Here, finding the solution to $D(d)$ implies finding the solution to $\Lambda$. In this case, with the help of equations~(\ref{eqn:rgx}) (the extended RG equations), it was possible to find the analytical expression for $\Lambda(d)$ in equation~(\ref{eqn:gamma_solution}). Notably, equations~(\ref{eqn:rgx}) stand on their own as a solution to the DLA dimensions, this is, independent of equation~(\ref{eqn:dphi}). Nevertheless, finding this solution was only possible because of the $\Lambda$ description (or ``$\Lambda$-space''), which allowed to define the region of validity for any solution of $D(d)$ through the KBW (Kesten-Ball-Wang) boundaries in equations~(\ref{eqn:kbw_conditions}) (see Fig.~3b). The values obtained for $D(d)$ from equations~(\ref{eqn:rgx}) not only are in good agreement with previously reported numerical and theoretical results (see Table~1), but they improve their accuracy for $d\geq 4$, and together with $D_B$ and $D_W$, they provide the exact solution ($D_B=d-1$) for any dimension larger than $d\approx 10$. It is only at $d\to\infty$ (specifically, $d\approx 100$) that all solutions become identical to $D_B$ (see Fig.~3). 

In this analysis it was considered that the mass-distribution of the DLA cluster is self-similar ($D_q=D$ for all $q$) \cite{sander2011,argoul1988,li1989,hanan2012,rodriguez2013}. However, even in the event of weak multifractality for $d=2$ and $d=3$ \cite{vicsek1990,kamer2013,rodriguez2013}, these results would still apply to the box-counting or Hausdorff dimension, $D_{q=0}$, and the correlation dimension, $D_{q=2}$, which are the most common dimensions reported for DLA; and even to the information-dimension, given the fact that for sufficiently large clusters, $D_{q=0}\approx D_{q=1}\approx D_{q=2}$ \cite{kamer2013,rodriguez2013}.

\textbf{\textsl{Solution to DBM.}} From equations~(\ref{eqn:dphi}) and (\ref{eqn:gamma_ansatz}), the DBM dimensions correspond to general case of $D(d,\eta)$. For fixed dimension, $d$, one has that $D(d,\eta)\to D(\eta)$, and $\Gamma(d,\eta)\to \Gamma(\eta)=\Gamma_0\eta^\chi$, with $\Gamma_0=d^{-1}\Lambda(d)$ given by equation~(\ref{eqn:gamma_solution}). Here, finding the solution to $D(\eta)$ implies finding the solution to $\chi$. In a first (numerical) approach, the values of $\chi$ for $d=2$ and $d=3$, were estimated using equation~(\ref{eqn:gamma_ansatz}) as a fitting-function (see Fig.~2a), yielding very good results (see Table 2). In a second (theoretical) approach, the hypothesis of maximum information-entropy production at the DLA point led to equation~(\ref{eqn:gamma_theory}), in which $\chi$ and $\Lambda$, become coupled. Together with equation~(\ref{eqn:dphi}), this result provides a theoretical solution to $D(d,\eta)$. 

From Table 2, this theoretical proposal for $D(\eta)$ is accurate for $d=2$ (within the statistical error of $\langle D(\eta)\rangle$ for $2\geq\eta\leq 3$), but it deviates significantly for $\eta>1$ in three dimensions (see Fig.~4a and 4b). A relevant reason behind this discrepancy could be associated to the small-size clusters used to measure the reported fractal dimensions in Table 2 \cite{tolman1989b,satpathy1986}, in which case, finite-size effects and crossovers can be accounted for the slow convergence of the fractal dimensions to their true values \cite{tolman1989,tolman1989b}. Another important reason behind this discrepancy could be associated to the fact that the functional form of $\Gamma(d,\eta)$ might not be exactly a power-law. We must recall that equation~(\ref{eqn:gamma_ansatz}) was introduced as an \textsl{ansatz}, from the evidence given by the data (see Fig. 2a). Hence, it remains as a quite good approximation as far as we can tell. Consequently, the functional form provided under the maximum information-entropy production hypothesis is expected to be different from equation~(\ref{eqn:gamma_theory}).

Similarly to DLA, in this analysis it was considered that the mass-distribution of the DBM clusters is self-similar \cite{sander2011,hastings2001}. Even in the event of weak multifractality, the same arguments that apply to the DLA clusters, apply to the DBM clusters. Even more, weak multifractality should only be expected for $0<\eta<4$, this is, sufficiently far away from the non-fractal limits \cite{hastings2001}. 

Lastly, equation~(\ref{eqn:dphi}) might resemble a functional form of the well-known Turkevich-Scher conjecture \cite{turkevich1985,turkevich1986}, $D=1+\alpha_{min}$, that relates the fractal dimension of the mass-distribution of the cluster, $D$, with the scaling of the maximum growth probability distribution defined on its surface, $\alpha_{min}$ (the minimum of an infinite set of exponents associated to scaling of the growth probability distribution, which is indeed a multifractal \cite{sander2011,jensen2003}). However, as originally formulated \cite{nicolas2017}, equation~(2) is neither dependent-on nor derived-from the Turkevich-Scher conjecture, and does not demonstrates its validity (a rigorous demonstration of the validity of this conjecture for $d\geq 2$ goes beyond the scope of this work). Hence, the results for the scaling of the mass-distribution of DLA and DBM clusters using equation~(2) are valid regardless of the Turkevich-Scher conjecture.  

\textbf{\textsl{Criticality and universality.}} The criticality of the DBM transition is understood in terms of $D(d,\eta)$ as a non-thermal order parameter. For example, for $d=2$, it has been suggested that the full collapse to linear clusters occurs at the ``critical'' value $\eta\approx 4$ \cite{hastings2001,halsey2002,mathiesen2002,mathiesen2008}. To address this point, the normalized dimension, $D^*=(D-1)/(d-1)$, can be used as the non-thermal order parameter of the system, where $D^*=1$, when $D=d$ at $\eta=0$, and $D^*=0$, when $D\to 1$ as $\eta\gg 1$. From equation~(\ref{eqn:dphi}), we have that $D^*=\exp(-\Gamma)$, is a continuous and monotonically decreasing function which obviously differs from the typical discontinuous power-law behaviour expected from the critical phase-transitions theory \cite{dimino1989c,gleiser2015}. Hence, the suggested ``critical'' point cannot be considered as such under this description. Nevertheless, it is still possible to define a \emph{transitional} point, $\eta_t$, i.e., a point where $D\approx 1$. Considering $D=1+\delta$, where $\delta\ll 1$, from equations~(\ref{eqn:dphi}) and (\ref{eqn:gamma_ansatz}), one has that $(\Lambda/d)\eta_t^\chi=-\log[\delta/(d-1)]$. For example, for $d=2$ and using $\chi=1.52$ (from theory), we have that $\eta_t(\delta=0.10)\approx 3.5$, or $\eta_t(\delta=0.05)\approx 4.1$, which are consistent with the reported value, $\eta\approx 4$ \cite{hastings2001}.

An interesting consequence of the previous results is that the fractal dimensions of the DBM can indeed be conformed to the universal description given by the normalized dimension, $D^*=\exp(-\Gamma)$, that as function of $\Gamma$ is independent of the initial cluster-configuration and the Euclidean dimension of the embedding space (see Fig.~4c). Under this description, the mean-field equation for $D_{MF}$ does not conform to the same curve, but it takes the form of its first-order approximation, $D^*=1/(1+\Gamma)$. Although this point might seem trivial, it implies that the mathematical formulation given by equation~(\ref{eqn:dphi}) is quite general, with the DBM being just one member of the family of morphological transitions that can be described under this framework \cite{nicolas2017}.


\section*{Final remarks} 

The main results of this work: the numerical-based analytical result for the DBM dimensions using equation~(\ref{eqn:gamma_ansatz}) as a fitting-function, the theory-consistent analytical expression for the dimensions of DLA in equations~(\ref{eqn:rgx}), and the theoretical result for the dimensions of the DBM using equation~(\ref{eqn:gamma_theory}); all of these results remain as good data- and theory-consistent approximations to the dimensions of both DLA and DBM as far as we can tell. In particular, a rigorous mathematical derivation of the information-function in equation~(\ref{eqn:gamma_ansatz}) is beyond the scope of this work. Nevertheless, these results provide one of the most simple analytical descriptions to most of the theoretical and numerical results for the fractal dimensions of the DBM and DLA reported in the literature. In addition, these results reveal an interesting universality regarding the mathematical description of the fractal dimensions of morphological transitions in terms of their information-function. We are confident that the characterization of these models, along with the data-set used in this analysis, will be useful to scientists working in fundamental and applied problems of complex growth phenomena.


\section*{Methods} 

\textbf{Construction of $\Phi(d)$.} In the RG model \cite{wang1989,wang1992}, the on-lattice DLA dimensions are given in discrete form by equations~(\ref{eqn:rg}), reprinted below, 
\begin{align}
D(k,d)&=1+\log \mu/\log 2,\nonumber \\ 
\mu(k,d)&=1 + 2\binom{d-1}{1}\phi_1 + \sum_{k=2}^{d-1}\binom{d-1}{k}\phi_k, \nonumber
\end{align}
where, $\mu=\mu(k,d)$ is inversely proportional to the maximum growth probability, $\sigma(d)_{max}=\mu^{-1}$, and where $\phi_k=\phi_k(d)$, with $k=1,2,\dots,d-1$, and $\phi_k=\phi_1,\phi_2,\dots,\phi_{d-1}$, are growth potentials that need to be determined for a given lattice configuration according to the relation $(d+1)\phi_k-k\phi_{k-1}-(d-k-1)\phi_{k+1}=1$. For example, for the square lattice, these relations lead to $\phi_1=1/3$ with $D\approx 1.74$; for a cubic lattice, they lead to $\phi_1=5/14$ and $\phi_2=3/7$, with $D\approx 2.52$. Independently of the lattice configuration, $\phi_\infty=1/2$, and $\mu_\infty=2^{d-2}+d/2$ are found in the $d\to\infty$ limit \cite{wang1992}. 

These lattice-\textsl{dependent} results are extended to the lattice-\textsl{independent} scenario by replacing the discrete lattice potentials, $\phi_k(d)$, with a continuous effective potential, $\Phi(d)$, this is, we consider that $\phi_k(d)\to\Phi(d)$ for all $k$. Then, by using the identity $\sum_{k=0}^{d-1}\binom{d-1}{k}=2^{d-1}$, equations~(\ref{eqn:rg}) can be rewritten in the forms presented in equations~(\ref{eqn:rgx}), reprinted below,
\begin{align*}
D(d)&=1+\log\mu/\log 2, \\ 
\mu(d)&=1+2(\mu_\infty-1)\Phi,\\
\mu_\infty(d) &=2^{d-2}+d/2.
\end{align*}

In the Supplementary Fig. 1S, we present a plot of $\Phi(d)$. In this plot, $\Phi(d)$ is shown as estimated from the data for $D$ in Table 1, using $\Phi[D(d)]=(2^{D-1}-1)/[2(\mu_\infty-1)]$. Similarly, by substituting $D_{MF}$ into $\Phi[D(d)]$ the mean-field approximation $\Phi_{MF}$ is also shown. In addition, the $D\geq D_K$ (Kesten), $D\geq D_B$ (Ball), and $D\leq D_W$ (Wang \cite{wang1992}) inequalities in equations~(\ref{eqn:kbw_conditions}), where, $D_K=(d+1)/2$, $D_B=d-1$, and $D_W=1+\log\mu_\infty/\log 2$ can also be plugged into $\Phi[D(d)]$ to establish the bounds ($\Phi_K$, $\Phi_B$, and $\Phi_W$) for any solution of $\Phi(d)$. 

The functional form of $\Phi$ is then constructed by taking into account that it must satisfy $\Phi=\phi_\infty=1/2$ for $d\to\infty$, whereas it must remain finite and positively defined as $d\to 1$ \cite{wang1992}, while maintaining a steady grow between limits. This suggest that $\Phi$ could be model using an ordinary Logistic equation of the form, $f(x)=L/(1+\exp[-R(x-x_0)])$. Here, $f(x)=\Phi(d)$, the initial dimension $x_0=d=1$, the ``carrying capacity'' or saturation limit $L=\Phi_\infty=1/2$, and steepness $R=r\log 2$, where the $\log 2$ factor is chosen to match the base of the original description and $r$ was determined by a numerical-fit to $\Phi(d)$ given by data. For this fit, only the numerical estimates of $D(d)$ for $d=2$ and $d=3$ were considered (which on average are the most reliable numerical results). From the previous, the general solution to the effective potential is given by,
\begin{align*}
\Phi(d)&=\Phi_\infty/[1+2^{-r(d-1)}],
\end{align*}
where $\Phi_\infty=1/2$, and $r=0.762\pm 0.014$. 

As shown in the Supplementary Fig. 1S, this solutions to $\Phi$ not only satisfies its expected asymptotic values but also, the necessary KBW-restrictions. Even more, in the same manner as the KBW bounds ($\Phi_K$, $\Phi_B$, and $\Phi_W$) for $\Phi$ were constructed by substituting $D_K$, $D_B$, and $D_W$, into $\Phi[D(d)]$, the corresponding bounds for the parameter $\Lambda(d)$, the maximum growth probability $\sigma_{max}(d)$, and the fractal dimensions $\Delta D$, can also be constructed by substituting $D_K$, $D_B$, and $D_W$, into their corresponding relations, 
\begin{align*}
\Phi[D(d)] &=(2^{D-1}-1)/[2(\mu_\infty-1)],\\
\Lambda[D(d)] &=-d\log[(D-1)/(d-1)],\\
\sigma_{max}[D(d)] &=\mu[D(d)]^{-1},\\
\Delta D(d) &=D-(d-1).
\end{align*}

\textbf{Alternative construction of $\Phi(d)$.} Another method to determine $\Phi$ is by finding its approximation as $d\to 2$. This is done by considering $\phi_k=1/2$ (the limit-value of $\phi_k$ as $d\to\infty$) for all $k\geq 2$ (this is $d\geq 3$) in equations~(\ref{eqn:rg}). This leads to $\mu(d)=\mu_\infty+(d-1)(2\phi-1)$, where $\phi_1\to\phi(d)$ is now a continuous function of $d$. We found that this $\phi(d)$ can be either determined by an adequate logistic function or heuristically constructed. Here we present the results for the latter. 

To construct $\phi(d)$, notice that for $d\to 2$ we have that $\mu(d\to 2)=1+2\phi$ together with $D=1+\log\mu/\log 2$, lead to $\phi(d\to 2)=(2^{D-1}-1)/2$. Here, $D(d\to 2)$ can be linearly approximated as $D(d\to 2)=1+(d-1)/\sqrt{2}$, where the slope of $1/\sqrt{2}$ is chosen \textsl{ad-hoc} according to a previous result \cite{ball1986}, in such a way that $D=1$ for $d=1$, $D=1+1/\sqrt{2}\approx 1.71$ for $d=2$, and $D=1+2/\sqrt{2}\approx 2.41$ for $d=3$; or according to $\Delta D=D-(d-1)$, we have $\Delta D=1$ for $d=1$, $\Delta D=0.71$ for $d=2$, and $\Delta D=0.41$ for $d=3$. However, given that $\phi(d\to 2)$ approximates $\Phi$ at the $d\to 2$ limit, we want the value of $\Delta D$ at $d=3$ to be as small as possible ($\Delta D\to 0$). This is because $\Delta D$ is nothing but a measure of the deviation of $D$ from Ball's limit $D_\infty=d-1$, and in this $\phi(d\to 2)$ approximation, the $d\to\infty$ limit has been applied to all $\phi_k=\phi_\infty=1/2$ for $k\geq 2$ ($d\geq 3$), then, these terms should not be contributing factors to $\phi(d\to 2)$. A relation that satisfies the previous condition is $D(d\to 2)=1+[(d-1)/2]^{1/2}$, which at $d=1$ gives $\Delta D=1$, at $d=2$ gives $\Delta D=1/\sqrt{2}\approx 0.71$, and at $d=3$ gives $\Delta D=0$. 

From the previous discussion, and by looking at the form of $\phi(d\to 2)=(2^{D-1}-1)/2$, and by using, $D(d\to 2)=1+[(d-1)/2]^{1/2}$, we propose the general expression, $\phi(d)=[2^{\sqrt{(d-1)/2}}-1]/d$ as an \textsl{ansatz}. To recover $\Phi$, let us observe that $\mu(d)=\mu_\infty+(d-1)(2\phi-1)$ and $\mu(d)=1+2(\mu_\infty-1)\Phi$ must be equivalent relations, consequently, $\Phi=1/2+(d-1)(2\phi-1)/[2(\mu_\infty-1)]$. Using this alternative potential $\Phi=\Phi_2$, all relevant quantities, $\Lambda_2$, $\sigma_2$, and $\Delta D_2$, can be recovered (see Supplementary Fig. 1S). The numerical estimates given by $D_2$ are shown in Table 1. Here it is important to remark that, even though this approach provides a parameter-free solution that satisfies all KBW-restrictions for $d\geq 2$, it violates the Kesten's bound in $1<d<2$ (this is better seen in its $\Lambda_2$ behaviour). Certainly this anomaly can be attributed to the imprecise nature of the \textsl{ansatz} for $\phi(d)$. Thus, this particular proposal as a general solution to the DLA dimensions must be taken with caution.


\begin{table*}[ht!]
\caption{\label{table1}\textbf{DLA dimensions.} (First section) Numerical estimates for $D(d)$. (Second section) Theoretical estimates for $D(d)$. (Third section) Average of numerical estimates, $\langle D(d)\rangle$, mean-field estimates, $D_{MF}(d)$, and the estimates obtained with equation~(\ref{eqn:rgx}), including $D_2(d)$ from Methods. Error in measurements is shown when available.}
\begin{tabular}{lllllllll}
Source & $d=2$	& $d=3$	& $d=4$ & $d=5$	& $d=6$	& $d=7$	& $d=8$\\
\hline
Rodriguez \& Sosa \cite{rodriguez2013} & $1.711\pm0.008$ & $2.51\pm0.01$ &  & 	& &	& \\
Meakin \cite{meakin1983a,meakin1983b}  & $1.71\pm 0.07$	& $2.50\pm 0.08$ 	& 	& 	& &	&\\
& $1.71\pm 0.05$	& $2.51\pm 0.06$	& $3.32\pm 0.10$	&	& 	&	&\\
& $1.70\pm 0.06$	& $2.53\pm 0.06$ & $3.31\pm 0.10$	& $4.20\pm 0.16$ & $\approx 5.35$	&	&\\
Tolman \& Meakin \cite{tolman1989}	& $1.715\pm 0.004$	& $2.495\pm 0.005$ 	& $\approx 3.40$	& $\approx 4.33$	& $\approx 5.40$	& $\approx 6.45$	& $\approx 7.50$ \\
\hline
Turkevich \& Scher \cite{turkevich1985} & $1.67$	&$2.46$ &  & 	& &	& \\
Erzan, \textit{et al.} \cite{erzan1995}  & $1.71$ & $2.54$ &  & 	& &	& \\
Halsey \cite{halsey1994}  & $1.66$ & $2.50$ & $3.40$ & $4.33$	& &	& \\
Hentschel \cite{hentschel1984}  & $1.75$ & $2.52$ & $3.38$ & $4.29$	& $5.0$ &	&\\
Wang \& Wang \cite{wang1989,wang1992} & $1.74$ & $2.52$ & $3.34$ & $4.22$	& $5.13$ &	$6.08$	& $7.05$ \\
\hline
$\langle D(d)\rangle$ & $1.71\pm 0.01$ & $2.51\pm 0.01$ & $3.34\pm 0.05$ & $4.27\pm 0.09$ & $5.38\pm 0.04$ &	 &  \\
$D(d)$, equation~(\ref{eqn:rgx}) & $1.70$ & $2.51$ & $3.36$ & $4.24$	& $5.15$ &	$6.09$	& $7.05$ \\
$D_2(d)$ & $1.71$ & $2.50$ & $3.32$ & $4.12$	& $5.11$ &	$6.07$	& $7.04$ \\
$D_{MF}(d)$ & $1.67$ & $2.50$ & $3.40$ & $4.33$	& $5.29$ &	$6.25$	& $7.22$ \\
\hline
\end{tabular}
\end{table*}

\begin{table*}[h!]
\caption{\textbf{DBM dimensions.} Average dimension, $\langle D(d,\eta)\rangle$, for $d=2$ (first section) and $d=3$ (second section), with the corresponding $D_{MF}(d,\eta)$, and the numerical (N), and theoretical (T) estimates for $D(d,\eta)$ of this work. Data marked with the asterisk are excluded from the corresponding averages due to known limitations in their measurements.}
\begin{tabular}{llllllll}
Data & $\eta=0.5$ & $\eta=1$ & $\eta=2$ & $\eta=3$ & $\eta=4$ & $\eta=5$ \\
\hline
Niemeyer, \textit{et al.} \cite{niemeyer1984}* & $1.89\pm 0.01$ & $1.75\pm 0.02$ & $1.6$ & & \\
Hayakawa, \textit{et al.} \cite{hayakawa1986}* & $1.79\pm 0.01$ & & $1.47\pm 0.03$ & & & \\
Pietronero, \textit{et al.} \cite{pietronero1988} & $1.92$ & $1.70$ & $1.43$ & & & \\
Somfai, \textit{et al.} \cite{somfai2004} & & $1.71$ & $1.42$ & $1.23$ & & \\
Tolman \& Meakin \cite{tolman1989b} & & & $1.408\pm 0.006$ & $1.292\pm 0.003$ & & \\
S\'anchez, \textit{et al.} \cite{sanchez1993}* & & $1.61	$ & $1.35$ & $1.22$ & $1.08$ & $1.04$ \\
Amitrano \cite{amitrano1989} & $1.86$ & $1.69$ & $1.43$ & $1.26$ & $1.16$ & $1.07$ \\
Hastings \cite{hastings2001} & & & $1.433$	&$1.263$	&$1.128$	&$1.068$ \\
& & & $1.426$	&$1.264$	&$1.090$	&$1.030$ \\
& & & $1.435$	&$1.262$	&$1.078$	&$1.025$ \\
& & & $1.452$	&$1.243$	&$1.071$	&$1.009$ \\
$\langle D(\eta)\rangle_{d=2}$ 	&$1.89\pm 0.05$	&$1.70\pm 0.01$	&$1.43\pm 0.02$	&$1.26\pm 0.02$	&$1.11\pm 0.04$	&$1.04\pm 0.03$\\
$D(\eta)_{d=2}$, $\Lambda\approx 0.68$, $\chi\approx 1.37$ (N) 		&$1.88$		&$1.71$		&$1.41$		&$1.21$		&$1.10$		&$1.04$ \\
$D(\eta)_{d=2}$, $\Lambda\approx 0.68$, $\chi\approx 1.52$ (T) 		&$1.89$		&$1.71$		&$1.37$		&$1.16$		&$1.06$		&$1.02$ \\
$D_{MF}(\eta)_{d=2}$, $\Lambda=1$, $\chi=1$ 		&$1.80$		&$1.67$		&$1.50$		&$1.40$		&$1.33$		&$1.29$ \\
\hline
Satpathy \cite{satpathy1986} & & $2.48\pm 0.06$ & $2.11\pm 0.06$ & $1.96\pm 0.08$ & $1.75\pm 0.06$ & \\
& & $2.54\pm 0.06$ & $2.09\pm 0.06$ & $1.84\pm 0.07$ & $1.79\pm 0.08$& \\
Tolman \& Meakin \cite{tolman1989b} & & & $2.134\pm 0.001$ & $1.895\pm 0.004$ & & \\
Vespignani \& Pietronero \cite{vespignani1993} & & $2.50\pm 0.10$ & $2.13\pm 0.10$ & $1.89\pm 0.10$ & & \\
& & $2.49$ & $2.17$ & $1.91$ & & \\
& & $2.54$ & $2.21$ & $1.92$ & & \\
$\langle D(\eta)\rangle_{d=3}$ 	&	& $2.51\pm 0.03$	&$2.14\pm 0.04$	&$1.90\pm 0.04$	&$1.77\pm 0.03$	&\\
$D(\eta)_{d=3}$, $\Lambda\approx 0.84$, $\chi\approx 0.91$ (N)  			&	&$2.51$		&$2.18$		&$1.93$		&$1.74$		&\\
$D(\eta)_{d=3}$, $\Lambda\approx 0.84$, $\chi\approx 1.39$ (T) 			&	&$2.51$		&$1.96$		&$1.55$		&$1.29$		&\\
$D_{MF}(\eta)_{d=3}$, $\Lambda=1$, $\chi=1$  		  	&	&$2.50$		&$2.20$		&$2.00$		&$1.86$		&\\
\hline
\end{tabular}
\end{table*}

\break


\section*{Acknowledgments}

Authors acknowledge partial financial support by CONACyT through the Soft Matter Network and VIEP-BUAP, grant CAREJ-EXC17-G.

\section*{Author contributions} 

J.R.N.C.\ collected and analysed the data from the literature, and produced all the figures. J.L.C.E.\ supervised the research. Both authors contributed in the discussion of the results and the preparation of the manuscript.

\section*{Additional information}

{\bf Competing interests:} The authors declare no competing interests.

\noindent 
{\bf Data availability:} All data generated or analysed during this study are included in this published article.

\end{document}